# Highly anisotropic spin transport in ultrathin black phosphorus

Luke Cording[1], Jiawei Liu[2], Jun You Tan[2], Kenji Watanabe[3], Takashi Taniguchi[4], Ahmet Avsar[1,2,5,6*], Barbaros Özyilmaz[2,5,6*]

[1] School of Mathematics, Statistics and Physics, Newcastle University, Newcastle upon Tyne, NE1 7RU, United Kingdom.
[2] Centre for Advanced 2D Materials, National University of Singapore, Singapore 117546,
[3] Research Center for Functional Materials, National Institute for Materials Science, Tsukuba 305-0044, Japan.
[4] International Center for Materials Nanoarchitectonics, National Institute for Materials Science, Tsukuba 305-0044, Japan.
[5] Department of Physics, National University of Singapore, Singapore 117542, Singapore.
[6] Department of Materials Science and Engineering, National University of Singapore, Singapore 117575, Singapore.

*Corresponding authors: aavsar@nus.edu.sg, barbaros@nus.edu.sg

**In anisotropic crystals, the direction-dependent effective mass of carriers can have a profound impact on spin transport dynamics. The puckered crystal structure of black phosphorus (BP) leads to direction-dependent charge transport and optical response, suggesting that it is an ideal system for studying anisotropic spin transport. To this end, we fabricate and characterize high mobility encapsulated ultrathin BP-based spin-valves in four-terminal geometry. Our measurements show that in-plane spin-lifetimes are strongly gate tunable and exceed one nanosecond. Through high out-of-plane magnetic fields, we observe a five-fold enhancement in the out-of-plane spin signal case compared to in-plane and estimate a colossal spin lifetime anisotropy of ~ 6. This finding is further confirmed by Oblique Hanle measurements. Additionally, we estimate an in-plane spin-lifetime anisotropy ratio of up to 1.8. Our observation of strongly anisotropic spin transport along three orthogonal axes in this pristine material could be exploited to realize directionally tunable spin transport.**

Two-dimensional black phosphorus (BP) has emerged as a promising semiconducting material for spintronics research due to the observation of high charge mobilities[1,2] along with its moderate atomic mass, which lead to long spin-lifetimes while still allowing spin transport to be influenced by spin-orbit coupling effects. BP has a centrosymmetric crystal structure, belonging to the D$_{2h}$ point group[3], so spin relaxation is expected to be governed by the Elliott-Yafet (EY) mechanism[4,5]. Recent measurements of spin transport in BP confirm that it exhibits the EY mechanism with nanosecond spin lifetimes, demonstrating the potential of this material for future spintronics research[6]. The semiconducting nature of BP results in a strongly gate dependent spin signal, giving an additional degree of freedom for the modulation of spin transport[6]. Additionally, the only stable isotope of phosphorus has half-integral nuclear spin, which distinguishes it from graphene and semiconductors such as silicon and germanium as a unique platform for studying rich spin-dependent physics.

BP has a highly anisotropic orthorhombic crystal structure comprising layers of covalently bonded phosphorus atoms in a honeycomb lattice. However, due to lone electron pairs pointing out-of-plane at 45°, the energy of each layer is minimized by puckering to form a unique periodic ridged structure[7]. The crystallographic axis following these ridges is termed the 'armchair' direction, and the axis orthogonal to them is termed the 'zigzag' direction[1]. Due to the anisotropy of its crystal structure, BP exhibits pronounced direction dependence in various physical processes. For example, charge transport measurements indicate a 1.8-fold difference between zigzag and armchair electron mobilities[8], and out-of-plane measurements in BP suggest a three-fold higher mean mobility for in-plane electron transport compared to out-of-plane[9,10]. High anisotropy is also noted in Raman spectra[12] and thermal conductivity measurements[11]. Anisotropic spin transport is expected in BP as its unique crystal structure results in direction-dependent effective mass, the expected source of anisotropic SOC. In the EY regime, a dependence of spin mixing on the direction of spin magnetic moment is predicted to cause different spin relaxation times for out-of-plane and in-plane spins[13].

Spin-lifetime anisotropy is defined as the ratio $\zeta \equiv \tau_{s,\perp}/\tau_{s,\parallel}$, where $\tau_{s,\perp}$ and $\tau_{s,\parallel}$ are the spin-lifetimes for spins directed out-of-plane and in-plane, respectively. This property has attracted considerable attention over the past decade, as it provides insight into the mechanisms governing spin transport, and the relative strength of spin-orbit effects that spins are exposed to. For pristine graphene, it is predicted that pure EY relaxation would lead to $\zeta = 0.5$; experimentally, both oblique Hanle measurements and high field Hanle measurements indicate an absence of spin-lifetime anisotropy, with impurities and random spin-orbit fields raising the anisotropy to $\zeta \approx 1$[14,15]. Strong spin-lifetime anisotropy has been realized in heterostructures of graphene and transition metal dichalcogenides (TMDCs), imprinted by proximity effects from the spin-valley coupling in the TMDC layer[16–18]. In these devices, spin-lifetimes of $\tau_{s,\parallel} = 3$ ps and $\tau_{s,\perp} = 30$ ps yield $\zeta = 10$[19]. Similarly strong anisotropy in bilayer graphene, arising from its Berry curvature, has also been observed at low temperatures by applying an electric field[20,21]. In silicon, Rashba SOC is induced by an out-of-plane electric field, resulting in ζ tunable between 1 and 0.75[22]. The strong spin-orbit field anisotropy and direction-dependent high electron mobility together with the expected hyperfine interactions make BP an exceptional material for studying such spin anisotropy.

In this work, we investigate spin-lifetime anisotropy in fully encapsulated ultrathin BP. Our heterostructure devices are composed of BP and boron nitride (BN) layers, as shown in figure 1(a). A ~ 5.5 nm BP ribbon served as the spin channel and was placed on a ~ 20 nm BN substrate, which promotes high carrier mobility by reducing scattering effects[23]. The heterostructure was then encapsulated by a ~ 1.1 nm thick (3 layer) BN crystal, which protects the spin channel from adsorption of airborne contaminants, which can degrade channel performance[24]. Additionally, this top BN layer acts as a tunnel barrier, increasing the efficiency of spin injection into the BP channel[25]. A BP ribbon was selected such that diffusion of injected spins would occur along the zigzag direction in spin-valve measurements. This choice was made as we found that the BP crystals tended to cleave in a ribbon shape, consistent with earlier studies[26]. The crystal orientation was confirmed by Raman spectroscopy (figure 1(c)). All heterostructure fabrication was performed in a glove-box environment and a final annealing step was carried out at 250 °C for 6 hours under vacuum conditions (~ $10^6$ Torr). Electron beam lithography was used to define contact regions, and Co/Au contacts of thickness 35 nm/5 nm were deposited over the top BN layer, shown in figure 1(a) inset. The main presented device has a channel width of 4 μm, with a separation of 1.9 μm between its injector and detector contacts.

Four-terminal spin-valve geometry, shown in figure 1(a), was used to generate a pure spin current in the channel and detect it. During spin-valve measurements, a magnetic field was swept in the plane of the BP along the easy axis of the ferromagnetic contacts. To determine the spin-lifetime anisotropy, we performed Hanle measurements by applying an out-of-plane field under two regimes. In the low field regime (B < 0.1 T), the device behavior for in-plane contact magnetization was probed. In the high field regime (B > 1 T), the behavior was probed for out-of-plane contact magnetization. This geometry, which has been previously applied to graphene spin-valves to determine spin-lifetime anisotropy, is suitable[19,20] for our system as spin transport in our devices occurs at high carrier concentrations with minimal non-linear magnetoresistive scaling (see Method and Supplementary Figures 1&2). Measurements were conducted at 1.5 K as a function of back-gate voltage ($V_{BG}$).

Prior to the spin-dependent measurements, the charge transport properties of the BP channel were characterized by probing the source-drain current $I_{SD}$ while sweeping the back-gate voltage $V_{BG}$ under fixed source-drain voltage $V_{SD}$. In two-terminal (2T) geometry, our device shows clear semiconducting field-effect behavior, as shown in figure 1(d), with a threshold voltage of +20 V for electron conduction and –10 V for hole conduction. Note that the device shows strong n-type behavior compared to the typical p-type response of BP due to dipoles formed at the BN/Co interface which change the position of the Fermi level, in agreement with our earlier works[6,27]. The device was further characterized by measuring four-terminal (4T) resistance as a function of back-gate voltage, given in figure 1(d) inset. From the 2T and 4T data, we extract the gate-dependent contact resistance, which shows a decrease from a value of 12.1 kΩ at $V_{BG}$ = 30 V to 10.7 kΩ at $V_{BG}$ = 62.5 V. Electron mobilities of up to ~ 200 cm$^2$V$^{-1}$s$^{-1}$ and ~ 4000 cm$^2$V$^{-1}$s$^{-1}$ were measured in the 2T and 4T geometries at 1.5 K, respectively. We note that the observation of highly non-linear, symmetric, and weakly temperature-dependent $I_{SD}$-$V_{SD}$ characteristics in our device (Figure 1e) suggests that tunnelling is the dominant charge injection mechanism and that 3-layer BN provides the ideal interface for spin injection into BP[6].

Now, we focus on spin transport measurements. To verify that the BP channel supported spin transport and extract critical spin transport parameters, a low field Hanle precession measurement was made by sweeping the device with an out-of-plane magnetic field while measuring the non-local resistance $R_{NL}$. Figure 2(a) shows that $R_{NL}$ peaks at zero-field, and falls to zero with increasing magnetic field magnitude, which is a characteristic feature of precessional spin dephasing. The parameters governing spin transport have been extracted by fitting the data to the solution of the one-dimensional Bloch equations, given by

$$R_{NL} \propto \int_0^\infty \frac{1}{\sqrt{4\pi D_{s,\parallel} t}} \exp\left(\frac{-L^2}{4 D_{s,\parallel} t}\right) \exp\left(\frac{-t}{\tau_{s,\parallel}}\right) \cos(\omega_L t)\, dt$$

where L is the channel length and the Larmor precession frequency is given by $\omega_L$. An in-plane spin relaxation time of $\tau_{s,\parallel}$ = ~ 1.1 ns was extracted at $V_{BG}$ = 52.5 V, along with an in-plane diffusion coefficient of $D_{s,\parallel}$ = ~ 0.014 m$^2$/s, yielding a spin relaxation length $\lambda_{s,\parallel} = \sqrt{D_{s,\parallel} \tau_{s,\parallel}}$ value of ~ 4.0 μm. These values are comparable to those previously reported in BP[6]. We also note a spin signal of $\Delta R_{HP} \approx 0.16$ Ω, given by the difference at zero field between $R_{NL}$ for parallel and antiparallel contact magnetization.

Further Hanle measurements were performed, this time varying $V_{BG}$. The extracted spin parameters have been plotted as a function of $V_{BG}$ in figure 2(b), exhibiting peaks in the 50 V-57.5 V range. A maximal spin diffusion length of 4.11 ± 0.24 µm is observed near $V_{BG}$ = 55 V, before decreasing and disappearing beyond $V_{BG}$ = 62.5 V. We also note a strongly gate dependent peak in spin signal, indicating an absence of spin transport outside of the 40 V-62.5 V range. This gate-controlled OFF/ON/OFF transition is unexpected, given that we observe high electron conductivity in our device for all gate voltages greater than $V_{BG}$ = 20 V. Computer simulations we have performed using the model posited by A. Fert and H. Jaffrès[28] suggest that this behavior could originate from a conductivity mismatch: the contact and channel conductivities are both gate dependent, but the channel conductivity scales more dramatically with $V_{BG}$. A maximal spin signal is expected when these conductivities intersect, and the efficiency of spin transport is expected to decrease away from this point, leading to a peak in spin signal as seen in our spin transport measurements (see Supplementary Figure 3).

Having demonstrated the optimal gate voltage for spin transport in the device being considered, we then probed the spin-lifetime anisotropy of BP by applying a strong magnetic field of magnitude up to B = 2 T at $V_{BG}$ = 55 V and 57.5 V. We intentionally selected this range of gate voltage as we observe a peak in the amplitude of spin signal and magnetoresistive effects are expected to be minimum at such high carrier concentrations ($n \sim 4\times10^{12}$ cm$^{-2}$)[15,29]. At this field strength, the magnetization of the contacts is forced out-of-plane for B > 1 T, leading to the injection of spins directed out-of-plane. The non-local resistance arising from this measurement scheme is given in figure 3(a). Standard spin precession is observed at low magnetic fields (|B| < 0.25 T), which is magnified and shown in figure 3(b). The Hanle curve resulting from in-plane spins has a peak at $R_{NL,\parallel} \approx 0.05$ Ω. Increasing the magnitude of the magnetic field beyond 0.25 T causes an increase in the measured value of non-local resistance for both parallel and antiparallel contact magnetization, up to $R_{NL,\perp} \approx 0.25$ Ω at –1 T, which is greater than the in-plane signal by a factor of five. Increasing the magnetic field magnitude beyond 1 T does not result in any further change to the non-local resistance because precessional dephasing does not occur for spins directed out-of-plane, along the magnetic field. We also note from the inset of figure 3(b) that the out-of-plane spin signal is positive for applied fields of –2 T and 2 T, this occurs because the relative orientation of spin magnetic moment and contact magnetization is parallel for both conditions.

For devices where the spin diffusion length exceeds the channel length, such as the one we have examined, the spin-lifetime anisotropy may be analytically estimated by making a first-order Taylor series approximation to the model governing spin signal magnitude (see Supplementary Section 4), giving a spin-lifetime anisotropy:

$$\zeta \approx \frac{D_{s,\parallel}}{D_{s,\perp}} \left( \frac{\frac{\sigma_\perp R_{NL,\perp}}{\sigma_\parallel R_{NL,\parallel}} + \frac{L}{\lambda_{s,\parallel}}}{1 + \frac{L}{\lambda_{s,\parallel}}} \right)^2$$

where $D_{s,\parallel}$ and $D_{s,\perp}$ are the in-plane and out-of-plane diffusion coefficients, and $\sigma_\parallel$ and $\sigma_\perp$ are the conductivities for in-plane and out-of-plane spins. From recent measurements in BP vertical field-effect transistors, we deduce a conductivity anisotropy ratio of ~ 3 between in-plane and out-of-plane directions[9]. This value is also supported by earlier works investigating charge transport anisotropy[10]. We employ this ratio along with our high-field Hanle measurement data to find spin-lifetime anisotropies of 5.0 ± 1.1 and 5.7 ± 0.6 for gate voltages of 55 V and 57.5 V, respectively. The uncertainty in these values is propagated from the uncertainty in the in-plane and out-of-plane spin signals, which in turn comes from noise in the measurements. From the Hanle components of the high field sweeps, in-plane spin-lifetimes of 1.15 ± 0.19 ns and 1.34 ± 0.11 ns have been extracted for 55 V and 57.5 V. From these data, we infer that out-of-plane spins have lifetimes of 5.8 ± 1.3 ns and 7.6 ± 0.8 ns for 55 V and 57.5 V.

To confirm the spin-based origin of the high field Hanle signal and the conclusions of these measurements, we performed measurements in the oblique Hanle scheme by finding the low field Hanle curve as a function of the angle of the applied field, β (see Supplementary Figure 4). The results of this are shown in figure 3(c). We find that the oblique Hanle ratio $R_{NL}/R^0_{NL}$ increases as a function of $\cos^2(\beta)$, which can only be explained by spin dependent behavior, indicating that the similar signal seen in the high field Hanle measurements originated from out-of-plane spin transport. Additionally, $R_{NL}/R^0_{NL}$ trends larger than the linear ratio for an isotropic system (given by the black dashed line), indicating that the spin lifetime anisotropy ratio is much greater than 1, in agreement with the high-field Hanle measurements. Furthermore, the oblique Hanle ratio signature for a spin lifetime anisotropy of 5.7 as was measured by our high field Hanle measurements (given by the red solid line) falls within the confidence interval for the oblique Hanle data.

To further characterize the spin channel and gain insight into the source of the strong spin-lifetime anisotropy, we performed spin-valve measurements while sweeping the device with an in-plane magnetic field. A representative scan taken at $V_{BG}$ = 50 V is shown in figure

4. Switching the relative direction of the detector and injector contact magnetization fields from parallel to antiparallel induces a discontinuous change in non-local resistance of $\Delta R_{SV} \approx 0.15$ $\Omega$ due to the presence of spin polarization at the detector. We also note a peak in the non-local resistance at zero field, a signal not generally seen for spin-valve measurements made with a field parallel to the injected spins. Our gate-voltage dependent spin-valve measurements suggest that this dephasing signal is present only across the range that our devices support spin transport, indicating that the signal directly results from spin dephasing and cannot be due to background magnetoresistive effects. To ensure that the dephasing is not an artefact of an extrinsic out-of-plane field component; the result for a similarly encapsulated graphene spin-valve taken in the same measurement set-up under comparable field geometry is shown in figure 4 inset, note the absence of spin dephasing around B = 0 T (see Supplementary Figure 5 for further discussion). The field dependence of this additional signal seen in BP is indicative of spin dephasing induced by spin-orbit effects in the out-of-plane direction, orthogonal to the spins in the channel.

Now we discuss the origin of this anisotropic spin transport. Previous instances of strong spin-lifetime anisotropy in TMDC heterostructures have been attributed to the protection of out-of-plane spins by spin-valley coupling[19–21]. However, BP demonstrates significantly smaller spin-valley splitting than TMDCs[30,31], making it unlikely to be the origin of the observed anisotropy. The presence of Rashba SOC, which explains the anisotropy in silicon, does not adequately justify the anisotropy seen here as it tends to preserve in-plane spins compared to those directed out-of-plane, resulting in anisotropy of $\zeta \lesssim 1$[14,15,22]. First-principles calculations assuming spin relaxation occurring purely by the Elliott-Yafet mechanism predict shorter spin-lifetimes for out-of-plane spins[13,30], meaning that the preservation of out-of-plane spins is likely to arise from a source outside of the Elliott-Yafet framework. It is also unlikely that the large anisotropy we observe originates from the hyperfine interaction as spin-valve measurements in materials with strong hyperfine interactions exhibit a characteristic antisymmetric dissimilar signal around B = 0 T[32]. Here, we note that a similar signature of spin dephasing to the one we note in our spin-valve measurements has been previously observed in germanium spin-valves, where it is attributed to g-factor anisotropy arising from its crystal structure[33], whereby fields directed along valley axes are exposed to a different g-factor to fields orthogonal to valley axes. This g-factor anisotropy transforms the Hamiltonian contribution from the applied magnetic field, causing the spins to feel an additional effective field component along a valley axis. If this new field component is not parallel to the spin magnetic moment, dephasing is induced. The presence of g-factor anisotropy in BP is predicted by theoretical work, arising due to interband coupling[34]. An anisotropic g-factor, if present in BP, could serve as a possible origin for the large out-of-plane spin-lifetime. The valley axis in the band structure of BP is located at the Z point of high symmetry, corresponding to momenta directed along the z-axis in real space, out-of-plane. Hence, out-of-plane spins would be subject to different g-factor compared to in-plane spins, increasing their spin-lifetime.

The small size of black phosphorus and boron nitride crystals that can currently be exfoliated precludes studying the anisotropy via spin transport measurements with diffusion along the orthogonal axes. Indeed, orthogonal Hanle precession measurements will yield little information about the spin-lifetime anisotropy, as the precessing spins are exposed to an average of spin-orbit coupling from the armchair and zigzag directions regardless of the diffusion direction. However, we can estimate the in-plane anisotropy by comparing the spin valve measurements to the low-field Hanle precession measurements. In the spin-valve measurements, the spins are directed along the armchair direction but diffuse along the less mobility zigzag direction, so the spin signal will exclusively be influenced by spin-orbit coupling along the armchair direction. Whereas, in the Hanle precession measurements, the field is out-of-plane, so spins precess in-plane direction exposing to SOC effects from an average of all in-plane directions. Therefore, the extracted spin parameters, and in turn the Hanle signal estimated using these parameters, will contain contributions from the zigzag and armchair directions. For measurements in an isotropic system such as graphene, the spin signals are demonstrated to be equal[15–17,19]. Here, we similarly observe comparable signals: spin-valve measurements gave a spin signal of $\Delta R_{SV} \approx 0.15$ Ω, whereas Hanle precession measurements resulted in a signal of $\Delta R_{HP} \approx 0.16$ Ω. Here, it is important to note that such comparable signals do not mean that spin transport in this system is also isotropic. We approximate spin anisotropy in the same manner as before, this time assuming that the in-plane spin-lifetime for Hanle measurements is the mean of lifetimes from spins exposed to SOC effects from the zigzag and armchair directions. We also take into account the anisotropic in-plane mobility of BP, which leads to a ratio of in-plane diffusion coefficients and conductivities of ~ 1.8 and 1.5, respectively for BP of similar thickness[8]. We estimate that spins in Hanle precession measurements experience a diffusion coefficient which is given by an average of the in-plane directions, which is a valid assumption given the monotonic change in conductivity between the armchair and zigzag directions with angle[1]. In doing so, we obtain the following equation (see Supplementary Section 4),

$$\frac{\tau_{s,A}}{\tau_{s,Z}} = \frac{4D_{s,Z}}{D_{s,Z} + D_{s,A}} \left( \frac{1 + \frac{L}{\lambda_{s,\parallel}}}{\frac{2\sigma_Z}{\sigma_Z + \sigma_A} \frac{\Delta R_{SV}}{\Delta R_{HP}} + \frac{L}{\lambda_{s,\parallel}}} \right)^2 - 1$$

giving a ratio of armchair to zigzag lifetimes of 1.11 ± 0.08 for VBG = 52.5 V, with uncertainty arising from noise in the measurements. These results show that the spin transport along the armchair direction has less diffusivity but longer spin lifetime than the zigzag direction. However, these contributions almost balance each other and result in comparable spin signal in armchair and zigzag directions. By following the same protocol, we estimate an in-plane anisotropy ratio of 1.02 ± 0.11 for $V_{BG}$ = 45 V, and 1.83 ± 0.29 for $V_{BG}$ = 55 V. Theoretical work predicts that the symmetry of BP obstructs Elliott-Yafet relaxation for SOC effects along the armchair direction, in agreement with our finding[13].

The strong spin-lifetime anisotropy we report adds to the growing body of evidence for the potential of BP in spintronics applications. This ability to directionally modulate spin-lifetime by a factor of up to 5.7 exemplifies the versatility of this novel material. Along with the electrically tunable carrier concentration arising from its semiconducting nature, BP could provide a platform for superlative control over spin transport. Additionally, this result indicates that BP is a fascinating candidate for developing heterostructures with TMDC layers. The proximity effects of the spin-valley coupling in the TMDC layer could potentially augment the intrinsically high anisotropy seen in BP and enable directional control of spin transport for future spintronics devices.

## ACKNOWLEDGEMENTS

The authors acknowledge helpful discussions with Jaroslav Fabian and Alberto Ciarrocchi. A. A. acknowledges support by the National Research Foundation, Prime Minister's Office, Singapore (NRFF14-2022-0083). B.Ö. acknowledges support by the National Research Foundation, Prime Minister's Office, Singapore, under its Competitive Research Program (CRP award no. NRF-CRP22-2019-8), the NRF Investigatorship (NRFI award no. NRF-NRFI2018-08), MOE-AcRF-Tier 2 (Grant No. MOE-T2EP50220-0017) and the Medium-Sized Centre Programme. K.W. and T.T. acknowledge support from the Elemental Strategy Initiative conducted by the MEXT, Japan, and the JSPS KAKENHI (Grant Numbers 21H05233 and 23H02052).

## AUTHOR CONTRIBUTIONS

A.A. and B.Ö. designed and coordinated the work. A.A. and J.Y.T. fabricated the samples. A.A. and J. Liu performed transport measurements. L.C. and A.A. performed simulations. K.W. and T.T. grew the hBN and BP crystals. L.C., A.A., and B.Ö. analyzed the results and wrote the manuscript with inputs from all authors.

## COMPETING FINANCIAL INTERESTS

The authors declare no competing financial interests.

## FIGURE CAPTIONS

**Figure 1 | Device fabrication and charge transport characterization. a,** Schematics of a non-local spin-valve geometry where the charge current path is isolated from the spin detection region. **b,** Optical image of a typical BP-based heterostructure prepared for spin-valve measurements. Heterostructure layers are outlined and highlighted. The thicknesses of the BN substrate, BP channel and the encapsulating BN layer are ∼ 20 nm, ∼ 5.5 nm and ∼ 1.1 nm, respectively. Inset shows the optical image of the device after the metallization process. **c,** Raman spectra of BP crystal. Blue and red lines represent measurements taken along 50° and 90°, respectively. Insets show the $\theta$ dependence of the $A_g^2 / A_g^1$ intensities. $\theta$ is shown in b. **d,** Back gate voltage ($V_{BG}$) dependence of bias current ($I_{SD}$) at fixed bias voltages ($V_{SD}$). Electron conduction begins at a threshold voltage $V_{TH} = +20$ V. Inset shows the $V_{BG}$ dependence of four-terminal, two-terminal and injector-detector contact resistances measured at 1.5 K. **e,** $V_{SD}$ dependence of $I_{SD}$ at fixed $V_{BG}$, measured at 1.5 K. Inset shows the $V_{SD}$ dependence of $I_{SD}$ at 300 K and 4 K for $V_{BG} = V_{TH} + 35$ V.

**Figure 2 | Nano-second spin lifetimes and gate-dependent spin transport. a,** Non-local signal as a function of out-of-plane magnetic field. Blue and red horizontal arrows represent the relative magnetization direction of the detector and injector contacts for each sweep. The maximum spin signal is measured to be $\Delta R_{HP} \approx 0.16$ Ω, given by the difference in non-local spin signal at zero field for measurements with contacts in parallel and antiparallel configuration. Measurement performed at $V_{BG} = 52.5$ V at 1.5 K. **b,** $V_{BG}$ dependence of spin-lifetime, diffusion coefficient, and spin diffusion length at 1.5 K. Error bars are calculated based on the standard errors resulting from the curve-fitting procedure applied to the Hanle data. These error bars are typically less than 3% of the mean values represented by the black-colored spheres, and they primarily stem from the noise present in the taken data.

**Figure 3 | Anisotropic spin relaxation. a-b,** Non-local signal ($R_{NL}$) as a function of the perpendicular magnetic field ($B_\perp$). Initially, the magnetization of the injector electrode is prepared parallel (red line) or antiparallel (blue line) to the magnetization of detector. In the low field range (**b** and grey shaded region in **a**), typical Hanle spin precession curves are obtained. Above a field strength of ~ 0.9 T, the magnetization directions of the injector and detector align along the out-of-plane direction (red shaded region in **a**). This allows the comparison of spin signals due to spins injected parallel ($\Delta R_{SV}^{\rightarrow}$) and out-of-plane ($\Delta R^{\uparrow}$) to the black phosphorus layer. The inset of **b** shows the data obtained while the magnetic field is swept from -2T to 2T. The magnitude of $\Delta R^{\uparrow}$ is independent of the field polarity. **c,** Ratio of non-local resistance measured when the out of plane spin component is dephased ($R_{NL}$) compared to the zero-field Hanle signal ($R^0_{NL}$) as a function of $\cos^2(\beta^*)$, where $\beta^*$ is the angle between the detector magnetization and the applied field. Also plotted are the theoretical lines for the case of an isotropic device (black dashed line), and the case of a spin lifetime anisotropy of 5.7 (red solid line), as was obtained through the high field Hanle measurements. The error bars are due to uncertainties in determining $R_{NL}$ and $R^0_{NL}$, arising from the measurement noise and zero-field spin dephasing effect as discussed in Supplementary Section 5. The mean values are represented by the black-colored spheres.

**Figure 4 | Indication of spin-dephasing near zero in-plane field.** Non-local signal ($R_{NL}$) as a function of in-plane magnetic field ($B_\parallel$). Red and blue directional arrows indicate the direction of magnetic field sweep. Measurement made at $V_{BG}$ = 50 V and 1.5 K. Discontinuous drop in signal of $\Delta R_{SV} \approx 0.15\ \Omega$ is seen upon changing the relative orientation of injector and detector contact magnetization directions. Peak in signal around $B_\parallel$ = 0 T suggests field dependent spin dephasing, which could indicate the presence of an out-of-plane SOC field. Inset shows sweep for an encapsulated graphene spin-valve fabricated under similar conditions in comparable field geometry. Flat signal indicates a lack of spin dephasing.

**REFERENCES**

1. Liu, H. *et al.* Phosphorene: An Unexplored 2D Semiconductor with a High Hole Mobility. *ACS Nano* **8**, 4033–4041 (2014).
2. Li, L. *et al.* Black phosphorus field-effect transistors. *Nat. Nanotechnol.* **9**, 372–377 (2014).
3. Ribeiro-Soares, J., Almeida, R. M., Cançado, L. G., Dresselhaus, M. S. & Jorio, A. Group theory for structural analysis and lattice vibrations in phosphorene systems. *Phys. Rev. B* **91**, 205421 (2015).

4. Kurpas, M., Gmitra, M. & Fabian, J. Spin-orbit coupling and spin relaxation in phosphorene: Intrinsic versus extrinsic effects. *Phys. Rev. B* **94**, 155423 (2016).

5. Kurpas, M., Gmitra, M. & Fabian, J. Spin properties of black phosphorus and phosphorene, and their prospects for spincalorics. *J. Phys. Appl. Phys.* **51**, 174001 (2018).

6. Avsar, A. *et al.* Gate-tunable black phosphorus spin valve with nanosecond spin lifetimes. *Nat. Phys.* **13**, 888–893 (2017).

7. Carvalho, A. *et al.* Phosphorene: from theory to applications. *Nat. Rev. Mater.* **1**, 1–16 (2016).

8. Xia, F., Wang, H. & Jia, Y. Rediscovering black phosphorus as an anisotropic layered material for optoelectronics and electronics. *Nat. Commun.* **5**, 4458 (2014).

9. Kang, J. *et al.* Probing Out-of-Plane Charge Transport in Black Phosphorus with Graphene-Contacted Vertical Field-Effect Transistors. *Nano Lett.* **16**, 2580–2585 (2016).

10. Akahama, Y., Endo, S. & Narita, S. Electrical Properties of Black Phosphorus Single Crystals. *J. Phys. Soc. Jpn.* **52**, 2148–2155 (1983).

11. Lee, S. *et al.* Anisotropic in-plane thermal conductivity of black phosphorus nanoribbons at temperatures higher than 100 K. *Nat. Commun.* **6**, 8573 (2015).

12. Ribeiro, H. B. *et al.* Unusual Angular Dependence of the Raman Response in Black Phosphorus. *ACS Nano* **9**, 4270–4276 (2015).

13. Li, P. & Appelbaum, I. Electrons and holes in phosphorene. *Phys. Rev. B* **90**, 115439 (2014).

14. Tombros, N. *et al.* Anisotropic Spin Relaxation in Graphene. *Phys. Rev. Lett.* **101**, 046601 (2008).

15. Raes, B. *et al.* Determination of the spin-lifetime anisotropy in graphene using oblique spin precession. *Nat. Commun.* **7**, 11444 (2016).

16. Sierra, J. F., Fabian, J., Kawakami, R. K., Roche, S. & Valenzuela, S. O. Van der Waals heterostructures for spintronics and opto-spintronics. *Nat. Nanotechnol.* **16**, 856–868 (2021).

17. Avsar, A. *et al.* Colloquium: Spintronics in graphene and other two-dimensional materials. *Rev. Mod. Phys.* **92**, 021003 (2020).

18. Garcia, J. H., Vila, M., Cummings, A. W. & Roche, S. Spin transport in graphene/transition metal dichalcogenide heterostructures. *Chem. Soc. Rev.* **47**, 3359–3379 (2018).

19. Benítez, L. A. *et al.* Strongly anisotropic spin relaxation in graphene–transition metal dichalcogenide heterostructures at room temperature. *Nat. Phys.* **14**, 303–308 (2018).

20. Leutenantsmeyer, J. C., Ingla-Aynés, J., Fabian, J. & van Wees, B. J. Observation of Spin-Valley-Coupling-Induced Large Spin-Lifetime Anisotropy in Bilayer Graphene. *Phys. Rev. Lett.* **121**, 127702 (2018).

21. Xu, J., Zhu, T., Luo, Y. K., Lu, Y.-M. & Kawakami, R. K. Strong and Tunable Spin-Lifetime Anisotropy in Dual-Gated Bilayer Graphene. *Phys. Rev. Lett.* **121**, 127703 (2018).

22. Lee, S. *et al.* Synthetic Rashba spin–orbit system using a silicon metal-oxide semiconductor. *Nat. Mater.* **20**, 1228–1232 (2021).

23. Long, G. *et al.* Achieving Ultrahigh Carrier Mobility in Two-Dimensional Hole Gas of Black Phosphorus. *Nano Lett.* **16**, 7768–7773 (2016).

24. Avsar, A. *et al.* Air-Stable Transport in Graphene-Contacted, Fully Encapsulated Ultrathin Black Phosphorus-Based Field-Effect Transistors. *ACS Nano* **9**, 4138–4145 (2015).

25. Yamaguchi, T. *et al.* Electrical Spin Injection into Graphene through Monolayer Hexagonal Boron Nitride. *Appl. Phys. Express* **6**, 073001 (2013).

26. Hsiao, Y., Chang, P.-Y., Fan, K.-L., Hsu, N.-C. & Lee, S.-C. Black phosphorus with a unique rectangular shape and its anisotropic properties. *AIP Adv.* **8**, 105216 (2018).

27. Avsar, A. *et al.* van der Waals Bonded Co/h-BN Contacts to Ultrathin Black Phosphorus Devices. *Nano Lett.* **17**, 5361–5367 (2017).

28. Fert, A. & Jaffrès, H. Conditions for efficient spin injection from a ferromagnetic metal into a semiconductor. *Phys. Rev. B* **64**, 184420 (2001).

29. Chen, X. *et al.* High-quality sandwiched black phosphorus heterostructure and its quantum oscillations. *Nat. Commun.* **6**, 7315 (2015).

30. Farzaneh, S. M. & Rakheja, S. Extrinsic spin-orbit coupling and spin relaxation in phosphorene. *Phys. Rev. B* **100**, 245429 (2019).

31. Fan, X., Singh, D. J. & Zheng, W. Valence Band Splitting on Multilayer MoS2: Mixing of Spin–Orbit Coupling and Interlayer Coupling. *J. Phys. Chem. Lett.* **7**, 2175–2181 (2016).

32. Chan, M. K. *et al.* Hyperfine interactions and spin transport in ferromagnet-semiconductor heterostructures. *Phys. Rev. B* **80**, 161206 (2009).

33. Li, P., Li, J., Qing, L., Dery, H. & Appelbaum, I. Anisotropy-Driven Spin Relaxation in Germanium. *Phys. Rev. Lett.* **111**, 257204 (2013).

34. Zhou, X. *et al.* Effective $g$ factor in black phosphorus thin films. *Phys. Rev. B* **95**, 045408 (2017).

# METHODS

**Growth of crystals**

BN single crystals were obtained by using the temperature gradient method under high pressure and high temperature. Typical growth conditions were 3 GPa and 1,500 °C for 120 h. A solvent of the Ba–BN system was used to obtain high-purity crystals. The source of BN crystals was heat treated so as to reduce oxygen impurities at 2,000 °C under a nitrogen atmosphere. The recovered BN crystals were treated with strong acid (hot aqua regia) to remove residual solvent and washed with diluted water to supply them for the exfoliation process. BP crystals were obtained by the melt growth process under high pressure. Typical growth conditions were 2 GPa and 1,200 °C with a slow cooling of 1 °C min$^{-1}$. The starting material of the BP crystal (5N) was encapsulated in a BN capsule under an Ar atmosphere. The recovered BP crystals were mechanically removed from the BN capsule and then used for the exfoliation process.

**Measurement**

Cryogenic measurements were performed in a closed-cycle Oxford Instruments cryomagnetic system with a base temperature of ~ 1.5K. During two-terminal charge transport measurements, drain currents were measured using a Keithley Sourcemeter 2450, and a Keithley Sourcemeter 2400 was used to apply bias through the SiO$_2$ gate dielectric. Four-terminal charge transport measurements were performed by using a lock-in amplifier (Stanford Research SR830) at very low-frequency (~ 13 Hz) while the device is highly conductive. For gate-dependent spin transport measurements, we utilized a Keithley-2400 current source to apply fixed current and a nanovoltmeter (Keithley 2182A) was used while detecting spin current.

In previous works studying graphene spin valves, spin precession measurements with a weak oblique magnetic field have been used as an alternative method of extracting the spin-lifetime anisotropy near the charge neutrality point[15]. The oblique measurement scheme is used because graphene exhibits large non-linear magnetoresistive scaling with B$^2$ at low carrier densities, making device characterization difficult in the typical high-field Hanle measurement scheme. In semiconducting BP, we investigate spin transport when the device is in a conductive state with high carrier concentration, and we also note that minimal non-linear magnetoresistive scaling is exhibited in our devices (see Supplementary Figure 1&2). Hence, the background can be easily removed to isolate the spin dependent signal, which is symmetric and saturates at $|B| = 2\,T$. In black phosphorus, we chose high field measurements as the primary method for extracting the spin lifetime anisotropy because the additional dephasing signal present in our devices with an in-plane field and the air sensitivity of black phosphorus make the sensitive and protracted oblique Hanle measurements less accurate. Therefore, we have primarily characterized the spin-lifetime anisotropy in BP via high field out-of-plane measurements.

**DATA AVAILABILITY**

The datasets generated during and/or analysed during this study are available from the corresponding author upon reasonable request.

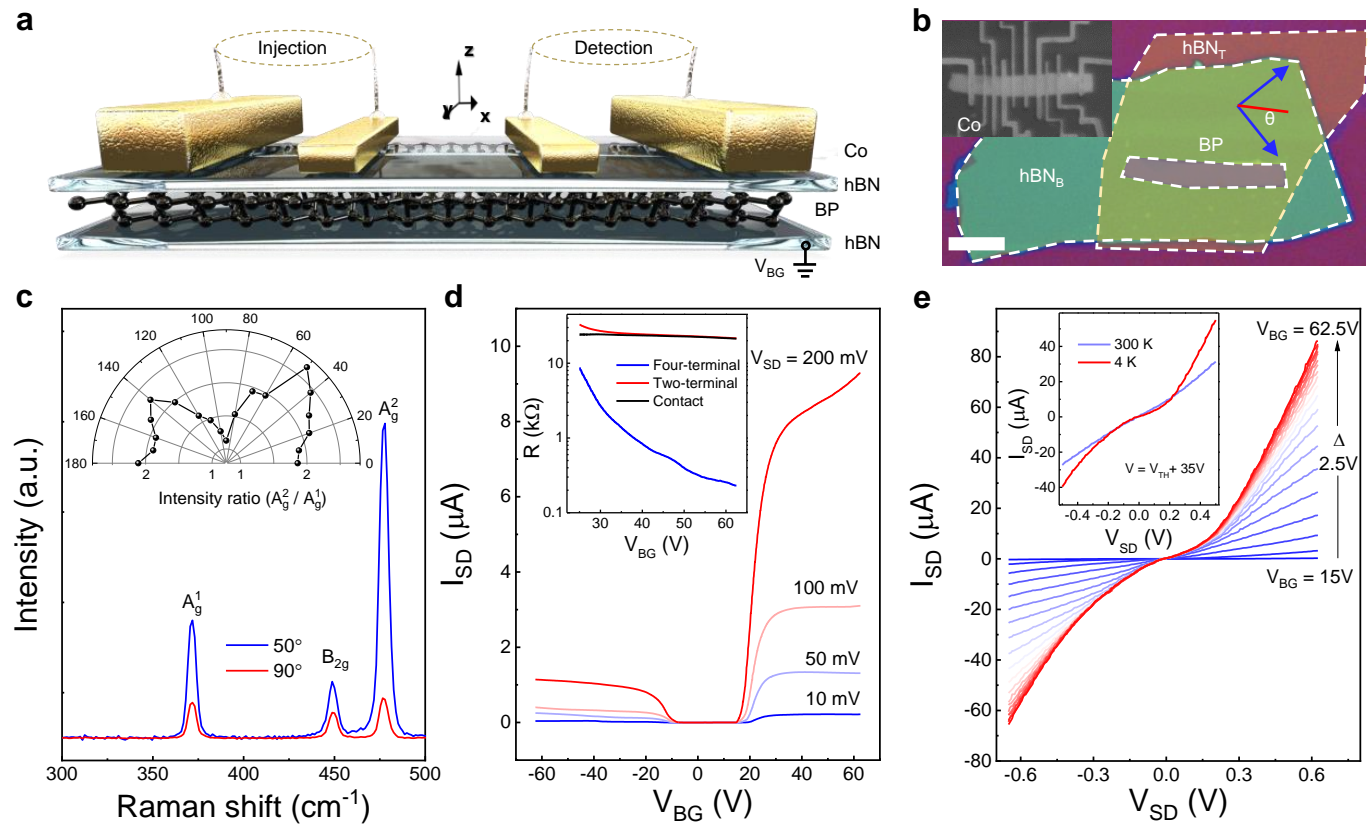

**Figure 1**

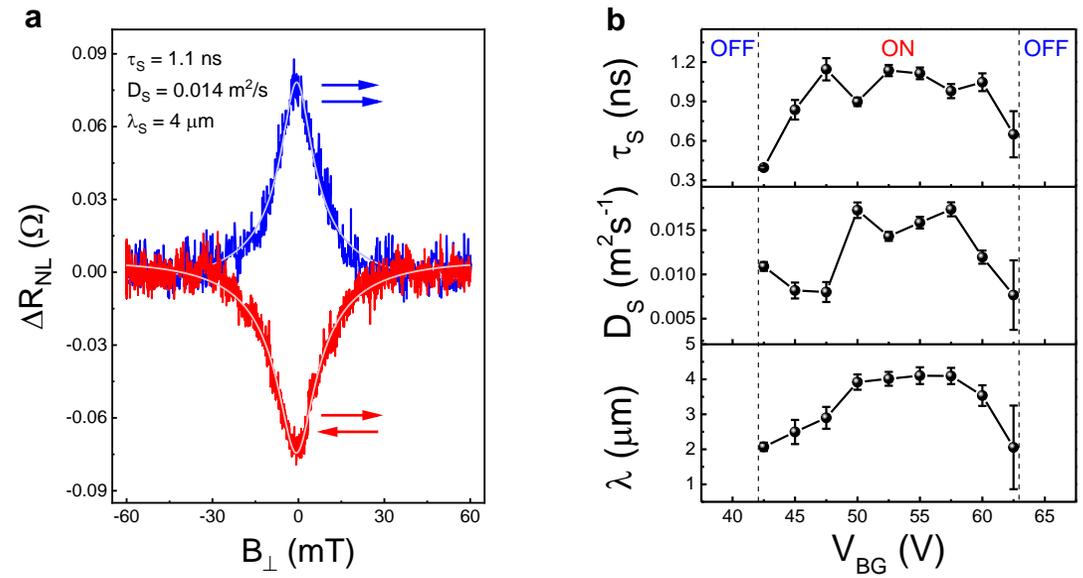

**Figure 2**

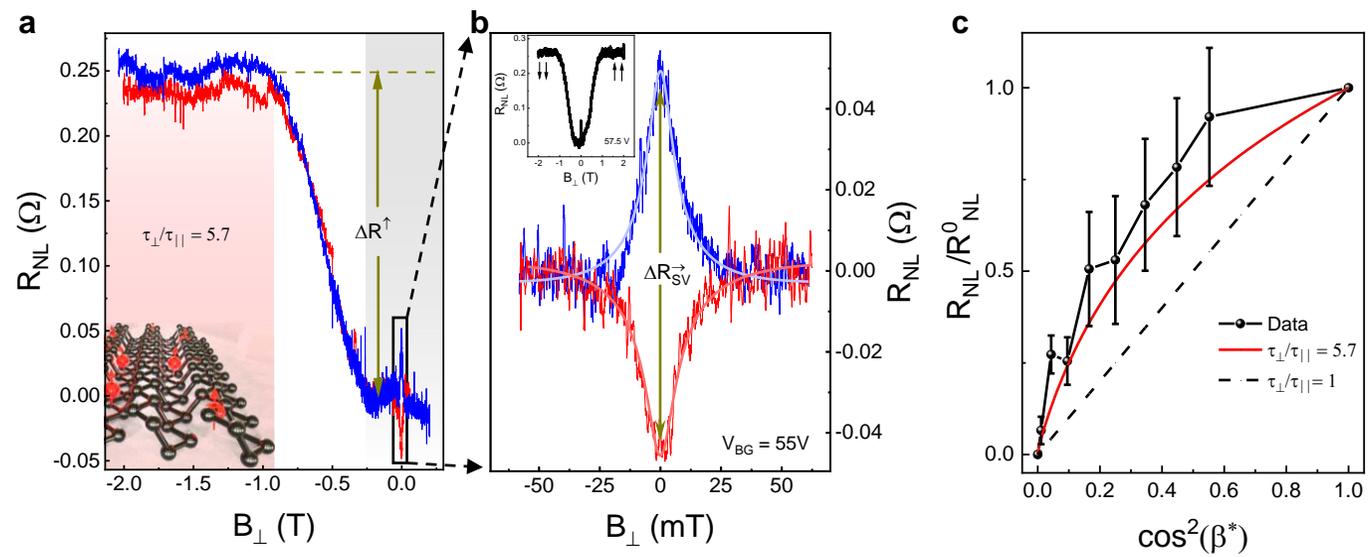

**Figure 3**

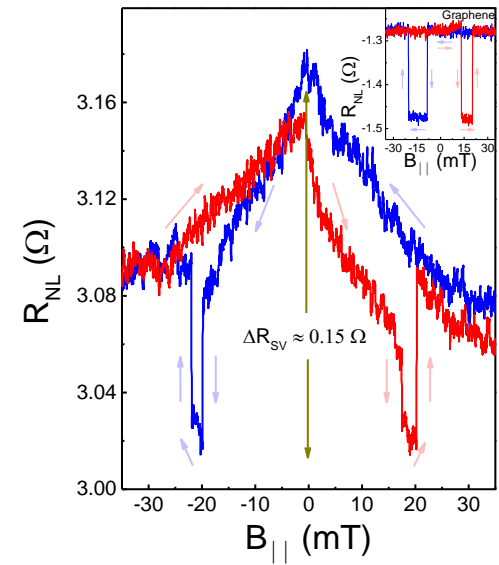

**Figure 4**